\documentclass{article}

\usepackage{spconf}
\usepackage{amsmath}
\usepackage{graphicx}
\usepackage{makecell}
\usepackage{url}
\usepackage{subfig}
\usepackage{float}

\title{Towards Leveraging End-of-Life Tools as an Asset: Value Co-Creation based on Deep Learning in the Machining Industry}
\twoauthors
  {Jannis Walk, Niklas K\"uhl}
	{Karlsruhe Institute of Technology\\
	\underline{\{walk, kuehl\}@kit.edu}}
  {Jonathan Sch\"afer}
	{CERATIZIT Austria GmbH\\
	\underline{jonathan.schaefer@ceratizit.com}}

\begin{document}
%
\maketitle
\begin{abstract}\itshape
Sustainability is the key concept in the management of products that reached their end-of-life. We propose that end-of-life products have---besides their value as recyclable assets---additional value for producer and consumer. We argue this is especially true for the machining industry, where we illustrate an automatic characterization of worn cutting tools to foster value co-creation between tool manufacturer and tool user (customer) in the future. 

In the work at hand, we present a deep-learning-based computer vision system for the automatic classification of worn tools regarding flank wear and chipping. The resulting Matthews Correlation Coefficient of 0.878 and 0.644 confirms the feasibility of our system based on the VGG-16 network and Gradient Boosting. Based on these first results we derive a research agenda which addresses the need for a more holistic tool characterization by semantic segmentation and assesses the perceived business impact and usability by different user groups.
\end{abstract}
%


\begin{figure*}[htbp]
    \centering
	\includegraphics[width=\linewidth]{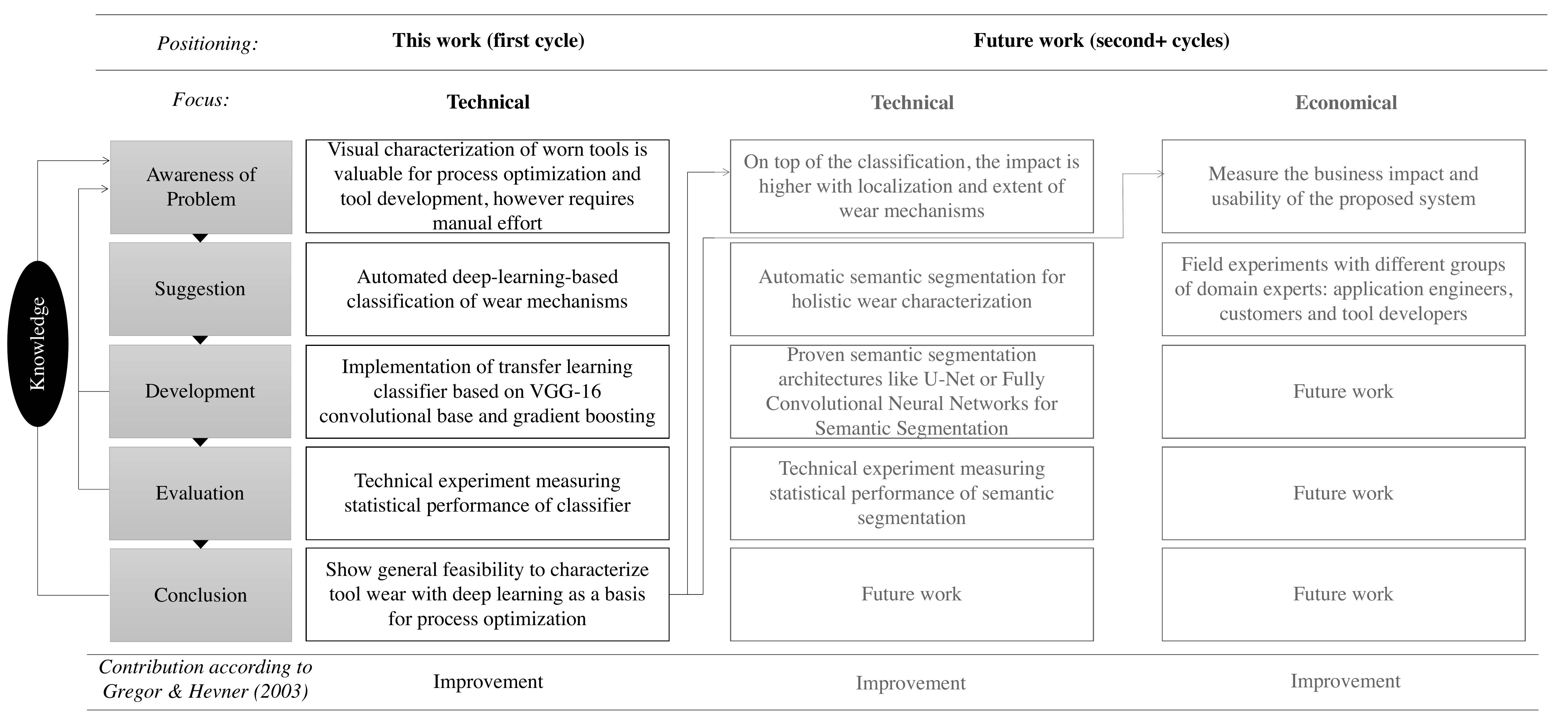}
	\caption{Overview of DSR cycles of the work at hand and the overall research endevour}
	\label{fig: DSR_cycle}  
\end{figure*}

\section{Introduction}

Sustainability is the key concept regarding the management of products having reached their end-of-life. Various approaches have been developed which suggest to implement sustainable end-of-life strategies already in the product development phase \cite{Gehin2008, Rose1999, Chan2007}. Such exemplary strategies range from refurbishing over remanufacturing to direct resale. 

We argue that products having reached their end-of-life have additional value, which exceeds the material value, for provider and customer. Thus, these products should be considered an asset. They can be leveraged to gain insights into their usage. This, in turn, can be utilized to positively impact earlier stages of the value chain through value co-creation which involves manufacturer and customers. 

Precisely, we propose to use worn tools from machining processes as a basis for easier and more objective optimization of customer's production processes. To this end, images of worn tools are automatically turned into valuable information by a deep-learning-based computer vision system. Information about occurrence, extent, and frequency of wear phenomena on the tools is usually the basis for understanding and improving machining processes. Due to the complexity of process optimization, tool manufacturers typically have dedicated teams of application engineers responsible for supporting the customers in the optimization of their processes. They often rely on the visual inspection of worn tools to understand potential problems in a machining process. This, however, is usually done manually with small and non-representative samples. Our proposed system enables the automatic characterization of a large quantity of worn tools. This leads to more reliable and information-rich results and thus facilitates an easier and more objective process optimization. To maximize the real-world impact, scalability and generalizability of our proposed system, we formulate the following requirements: 

Labelled training data should be the only required human input. As a consequence, the system can easily be trained for other tools or wear mechanisms. Also, the images for the testing and development of the system should be from real production processes.

In addition to enhancing process optimization, the insights based on our proposed system can also support the development of new tools. First, the development process itself can be accelerated since wear characterization is a frequent task in tool development and executed manually so far. Second, and more important, our system enables profound insights into potential problems of certain tools. So far, testing is mainly done internally and with standardized, simplified processes. With our proposed system it will be possible to analyze the wear mechanisms on a large quantity of tools used by customers in different real-world processes. This supports identifying promising directions for the development of new tools in the machining industry. 

The remainder of this work is structured as follows: in section \ref{researchDesign}, we present our research design. Subsequently, related work from different domains is introduced in section \ref{Related  Work}. Based on this we then present our first, already completed, design cycle in detail in section \ref{FirstCycle}. In section \ref{FutureCycles} we then present our agenda for future research. Afterwards, in section \ref{conclusion}, we summarize our work and describe limitations.

\section{Research Design}
\label{researchDesign}
As an overall research design, we choose Design Science Research (DSR), as it allows to consider the theoretical and practical tasks necessary when designing IT artifacts \cite{march1995design} and has proven to be an important and legitimate paradigm in information systems research \cite{gregor2013positioning}. For the design of the artifact, we follow the DSR process methodology and its individual phases according to Kuechler and Vaishnavi (2008) \cite{kuechler2008theory}, as we favor a clear differentiation between an abstract ``suggestion'' and a concrete, more programming-specific ``development''. The work at hand presents the first DSR cycle as part of a larger research endeavor. Our overall goal is to assess the following general research question:\\ \textbf{How can we utilize 
end-of-life tools to improve processes at the interface of tool manufacturer and customer?} 
In the work at hand, we complete the first cycle with the individual phases as illustrated in figure \ref{fig: DSR_cycle}. 
\noindent
We ask the following specific research question:  \\ \textbf{How can we design a system for deep-learning-based computer vision to automatically classify worn tools regarding their wear phenomena?} This research question forms the basis of our overall research endeavor and allows to draw conclusions regarding the future steps of our research project.

In terms of knowledge contribution, the presented work of the first cycle depicts an ``improvement'' according to Gregor and Hevner (2013) \cite{gregor2013positioning}, since we apply a novel method, i.e., supervised machine learning with deep neural networks \cite{LeCun2015}, to the existing problem of worn tool classification. In order to evaluate the resulting artifact, we use a technical experiment as proposed by Peffers, Rothenberger, Tuunanen and Vaezi (2012) \cite{peffers2012design}. We evaluate the statistical classification performances of the identified models. Figure \ref{fig: DSR_cycle} presents the activities of this first DSR cycle, as well as the future research activities containing additional two cycles, separated into the steps of problem awareness, suggestion, development, evaluation and conclusion. After elaborating on the state of the art of relevant fields for the research at hand in a designated rigor cycle \cite{hevner2007three}, we present all aforementioned steps for the first design cycle. Subsequently, we describe our research agenda for the second and third design cycle in section \ref{FutureCycles}.

\section{Rigor Cycle and Related Work}
\label{Related  Work}
To set a foundation for the remainder of this work, we review relevant literature from the body of knowledge. Several fields of research are of relevance, which we elaborate on in the following subsections: machining and wear mechanisms, deep learning and computer vision as well as value co-creation.

\subsection{Machining and wear mechanisms}
\vspace{-5pt}
Machining is ``one of the most important of the basic manufacturing processes'' \cite[p. VI]{Black1995}. It is applied in a variety of industries like aerospace, automotive, and the electro and energy industry. In general, machining describes the process of removing unwanted material from a workpiece \cite{Black1995}. The removal of unwanted material is generated by a relative motion between the cutting tool and the workpiece \cite{Boothroyd1989}. In regards to the different types of material, metallic workpieces are most widespread \cite{Black1995}. The tools used for machining can be regarded as consumables, as the occurrence of wear which ultimately results in a tool that can not be used anymore is inherent. For the first design cycle, we aim to show the general feasibility of our proposed system, therefore, we concentrate on the two main wear mechanisms we observed in our data set: flank wear (82.56\%) and chipping (55.40\%) of the cutting edge. We will briefly describe those in the following.

\begin{figure}[H]
    \centering
	\includegraphics[trim={3cm 3cm 3cm 3cm}, clip,width=0.8\linewidth]{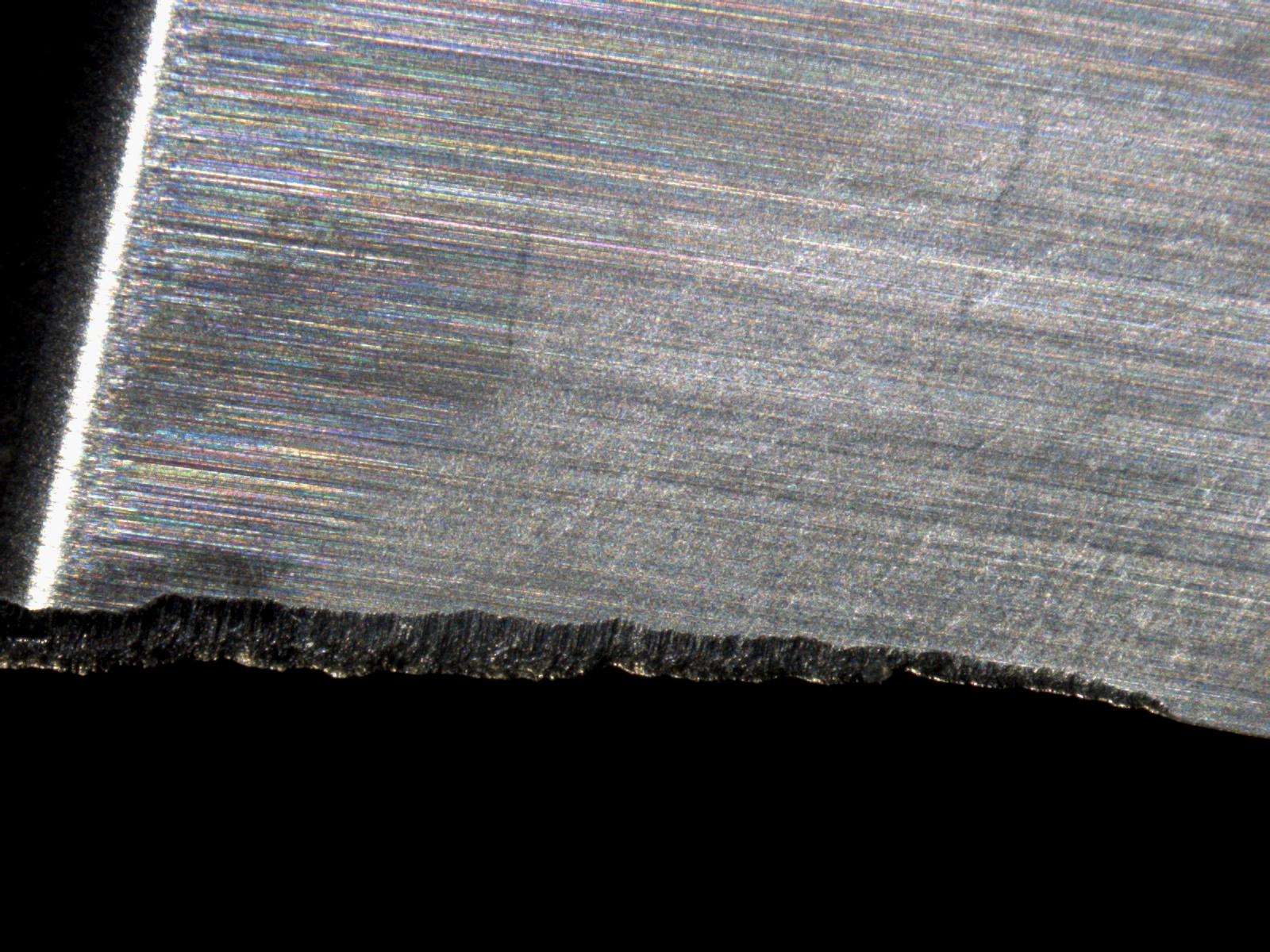}
	\caption{Example of flank wear.}
	\label{fig: VB_example}
\end{figure}

\begin{figure}[thb]
    \centering
	\includegraphics[trim={3cm 3cm 3cm 3cm}, clip,width=0.8\linewidth]{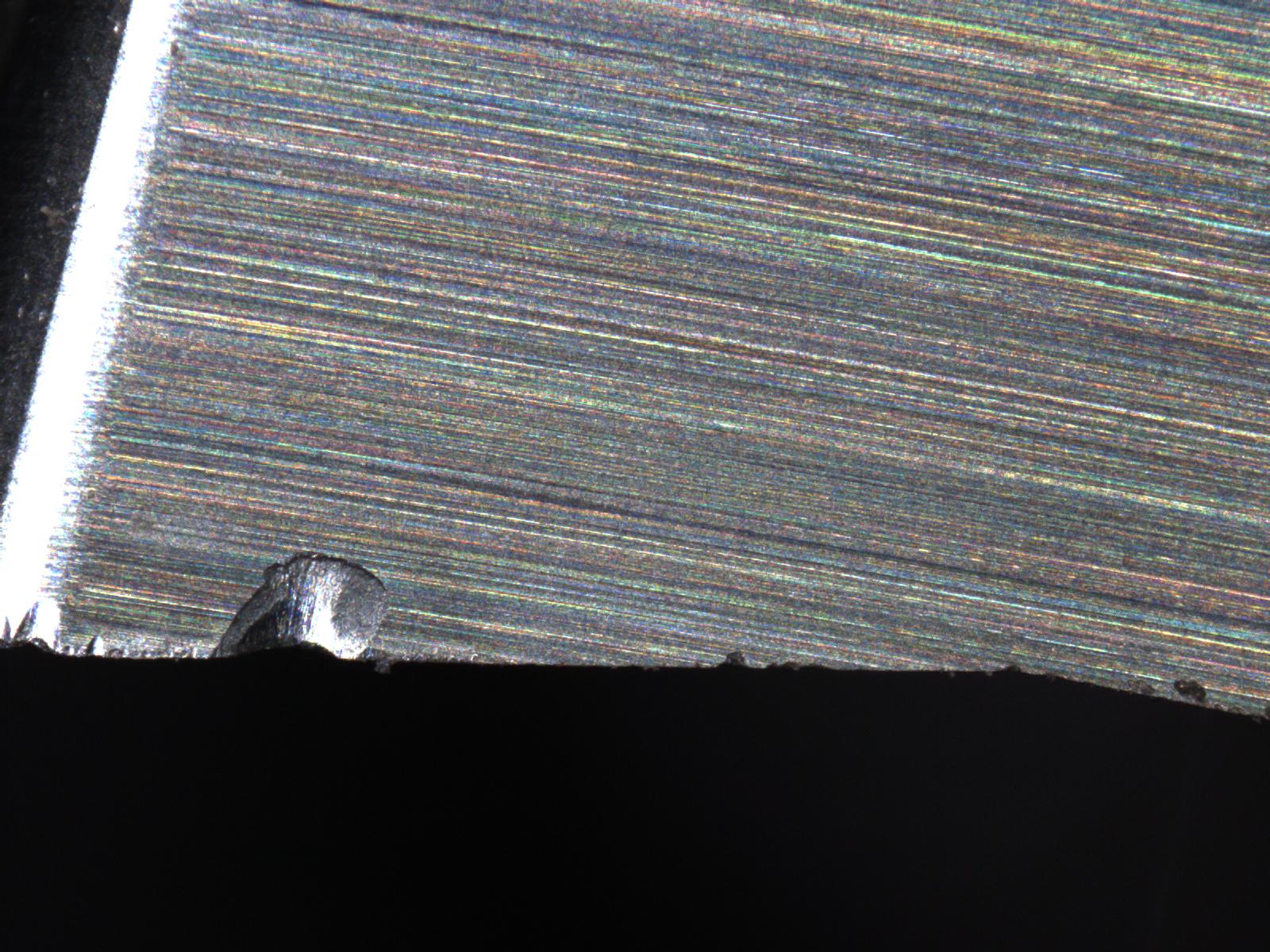}
	\caption{Example of chipping.}
	\label{fig: Chipping_example}       
\end{figure}

\textit{Flank wear} occurs due to friction between tool flank surface and workpiece \cite{Altintas2012}. It is unavoidable and thus the most commonly observed wear mechanism \cite{Siddhpura2013}.  As a consequence it is regarded as good criterion for tool-life, i.e. for deciding when to change a tool \cite{ISO3685}. An exemplary image with flank wear is depicted in figure \ref{fig: VB_example}.
\textit{Chipping} refers to particles of the cutting edge breaking off and thermal cracking \cite{ISO3685}. This is less common and also less desirable since it suddenly deforms the cutting edge and leads to poor surface quality on the workpiece. Figure \ref{fig: Chipping_example} shows an exemplary image with chipping.

For the research at hand the application of image processing techniques for tool condition monitoring is the most related field of machining research. Tool condition monitoring based on image processing techniques means that an automatic visual inspection is used to determine the wear state of cutting tools. This enables to decide whether a tool can still be used or not. In the following we briefly present research from this field.

Dutta et al. (2013) \cite{Dutta2013} provide a comprehensive review of the field of wear classification and measurement based on image processing. We briefly describe the papers most relevant for our research:
First, there is a multitude of research developing approaches for automatic wear measurement. Several articles describe systems for flank wear measurement for drills, which are based on traditional computer vision approaches \cite{Liang2007, Su2006, Duan2010}. Traditional computer vision refers to approaches like texture-based image segmentation and edge detection for which the user needs to fine-tune a multitude of parameters \cite{OMahony2019}. Another common approach is to classify the extent of wear into different classes. For example, Alegre et al. (2009) \cite{Alegre2009} use traditional computer vision algorithms (preprocessing like filtering and then automatic segmentation) to classify the flank wear on cutting inserts into low and high. Castej{\'{o}}n et al. (2007) \cite{Castejon2007} extract geometrical descriptors with traditional computer vision approaches and then use machine learning to classify if the wear on a given image is low, medium or high. Another stream of research works on classifying which wear mechanisms are visible on a given image. For instance, Schmitt et al. (2012) \cite{Schmitt2012} use traditional computer vision features (image statistics, surface texture, Canny analysis, histogram and Fourier coefficients) as input for a neural network which decides if the wear mechanism on the image is flank wear or tool breakage. Subsequently, they also apply an active contour algorithm to extract the wear region. In addition to the wear region, they compute the maximum and average wear perpendicular to the cutting edge. Lanzetta (2001) \cite{Lanzetta2001} proposes a system to detect all types of wear on cutting inserts. Depending on the concrete tool, several parameters have to be chosen by the user of the system.
Interestingly, this is the only identified article stating that the images are from cutting tools that were used in a real production environment---other articles either describe how they used the tools in their laboratory or do not elaborate on the environment.

Overall, we conclude that extant literature provides meaningful ideas for the development of an automated tool characterization system. However, none of the regarded research described above satisfies the requirements we formulated for our system in the introduction. Namely, that only labelled images are necessary as human input and that all images for testing and development of the system should be from real production processes. Furthermore, critically viewed, the performance of the systems developed in the articles above is often intransparent and hard to reproduce. Most data sets are rather small, e.g. Schmitt et al. (2012) \cite{Schmitt2012} use 15 images for the training of their neural network and 25 for testing. Also, they do not describe their data set in detail---it is unclear on how many images which wear mechanisms are visible. Thus, it is not clear if they also worked with images where more than one wear mechanism is visible. Other papers rely on a purely visual evaluation based on concrete image examples \cite{Liang2007, Duan2010, Lanzetta2001}).

\subsection{Deep learning and computer vision}
\label{subsec: DL and CV}
\vspace{-5pt}
Some of the systems for wear measurement and classification we just presented already use machine learning. However, they all rely on traditional computer vision approaches like edge detection to extract features from the raw images \cite{Alegre2009, Castejon2007, Schmitt2012}. With deep learning algorithms this becomes obsolete. Deep learning algorithms implement representation learning, i.e. they are able to directly process raw data and learn the relevant features for the task themselves \cite{LeCun2015}.
Even more importantly, deep learning algorithms have been proven to achieve far better results than the previous state-of-the-art techniques in many computer vision applications \cite{Krizhevsky2012, Voulodimos2018}. 
Specifically, convolutional neural networks are applied for computer vision tasks. The first and main part of these networks consists of a series of convolutional and pooling layers \cite{LeCun2015}.  In the convolutional layers filters are applied. These filters are learned from the data by backpropagation. Pooling layers ``merge semantically similar features into one'' \cite[p. 439]{LeCun2015}, a typical application is to compute the maximum over e.g., nine pixels. 
In a given layer, the respective operations (convolution or pooling) are applied to all inputs from the previous layer. This drastically reduces the amount of weights to be learned compared to fully-connected networks where the weight is distinct for each connection of two neurons. 
Depending on the concrete computer vision application, the output is computed directly by a convolutional layer \cite{Ronneberger2015} or by a series of fully-connected layers \cite{Krizhevsky2012}.

\subsection{Value Co-Creation}
\vspace{-5pt}
With the relevant research from a technical perspective at hand, we now regard related work from a business perspective.
Especially in the machining industry, the understanding of value has been mainly influenced by the goods-dominant logic: value is created (manufactured) by one firm and distributed in the market, usually through exchange of goods and money \cite[p. 146]{Vargo2008}. Other industries like the software industry, in contrast, have already adopted the idea of service-dominant logic where ``the roles of  producers and consumers  are not distinct, meaning that value is always co-created, jointly  and reciprocally, in interactions among providers and beneficiaries through the integration of resources and application of competences'' \cite[p. 146]{Vargo2008}. 

Several studies show that this value co-creation can be beneficial. For instance Nike, formerly also a product-centric company, successfully used a social networking site for co-creation with their customers. Among other benefits they also use the social networking site to learn about their customers' needs and preferences. Overall, they used the Internet engagement platform ``to establish customer relationships on a scale and scope as never before'' \cite[p. 10]{Ramaswamy2008} .
On a more general level, Kale et al. (2009) \cite{Kale2009} show that partnerships between companies generally help increasing firm value. 
In the remainder of this work, we take the perspective of service-dominant logic as well, as our general research question refers to the creation of value at the interface between provider and customer.\\ 
The subfield of "reverse use of customer data" is even more closely related to our research. Saarij\"arvi et al. (2014) \cite{Saarijarvi2014} describe three cases how customer data can be turned into information that directly supports customers' value creation. We build on this research and extend it since the cases of Saarij\"arvi et al. (2014) \cite{Saarijarvi2014} rely on usage data as a basis for value creation. We, however rely on products without any usage data. In that sense our analysis is forensic. We do not have access to any usage data and can only rely on the tool having reached its end-of-life and the observations we can make directly from it.
\subsection{Summary and Delineation}
Based on the related work described above we believe we can contribute to the body of knowledge on different levels: Our proposed system addresses the lack of reproducibility and generalizability in existing research on automatic wear characterization and measurement based on image processing. First, we aim to ensure reproducibility by a detailed description of both our data sets and the computer vision systems. To the same end, we will use acknowledged machine learning evaluation techniques. 

In regards to the missing generalizability, which we encountered in existing literature, we aim to utilize flexible, modern approaches. Existing research is based on traditional computer vision approaches, thus, a multitude of parameters need to be fine-tuned by the user. The recent developments in the area of deep learning facilitate end-to-end learning. Consequently, labelled training data is the only required human input for our proposed system based on deep learning. Thus, the system can be trained for other cutting tools or wear mechanisms without the need to fine-tune parameters.

From a business perspective our research contributes to the field of value co-creation and reverse use of customer data since it shows that these value creation mechanisms are also feasible based on forensic analyses.

\section{First Design Cycle: Wear mechanism classification based on Deep Learning}
\label{FirstCycle}
So far, we completed the first cycle of our research endeavor, which we present in this chapter.
\subsection{Awareness of Problem and Data Set}
Visual characterization of worn tools is an essential part of the optimization of machining processes. We conducted interviews with domain experts to better understand their general approach for this optimization. Usually, visual characterization is done manually. The necessary effort leads to small and non-representative samples of worn tools. Due to the advances in the deep learning field described in subsection \ref{subsec: DL and CV}, it seems plausible to apply deep learning for characterizing images of worn tools. Therefore, in the first cycle, we assess the feasibility of characterizing worn tools with deep learning. To be precise, we implement and evaluate two classification models: one for each of the two most prevalent wear mechanisms. We consider this a reasonable feasibility study since it gives an indication if and how deep learning algorithms are able to extract relevant features directly from the images in our data set.

 \begin{figure}[]
    \centering
	\includegraphics[trim={3cm 3cm 3cm 3cm}, clip,width=0.9\linewidth]{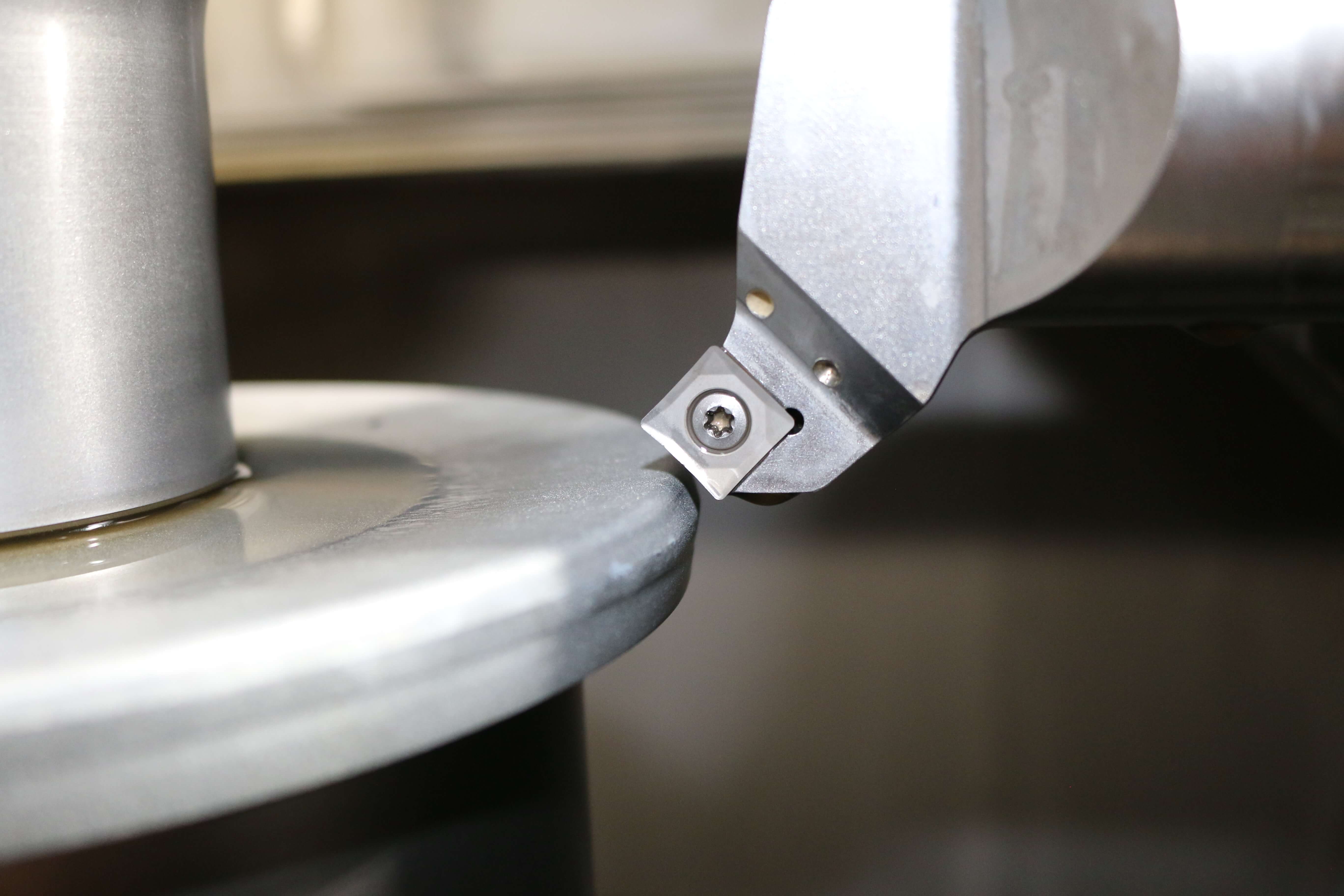}
	\caption{One of the two production processes.}
	\label{fig: productionProcess}       
\end{figure}

 Our data set consists of 648 images of worn cutting inserts from real production processes on two different machines. The type of cutting insert is always the same for this first design cycle. Figure \ref{fig: productionProcess} shows one of the two production processes. The workpiece to the left rotates at high speed such that the cutting insert to the right removes unwanted material; during production the workpiece and cutting insert are in direct contact. The images in our dataset show the flank side, i.e. the back side of the cutting edge.
 
To train and evaluate a classification algorithm we labelled the images manually. The first 60 images were labelled jointly by three domain experts. Afterwards labels were assigned individually, unclear cases were discussed by the three domain experts.
 
 Several wear mechanisms are present on the images. Table \ref{Table: FrequencyOfWearMechanisms} shows the absolute frequency of different wear mechanisms. A cutting edge could show no wear, if e.g., wear on other parts of the cutting insert prevent a utilization. Note, that the data implies the presence of more than one wear mechanism on many pictures. Due to the scarcity of data for all other wear mechanisms, we only consider flank wear and chipping as wear mechanisms for our first design cycle.
 
\begin{figure}[thb]
    \centering
	\includegraphics[trim={3cm 3cm 3cm 3cm}, clip,width=0.8\linewidth]{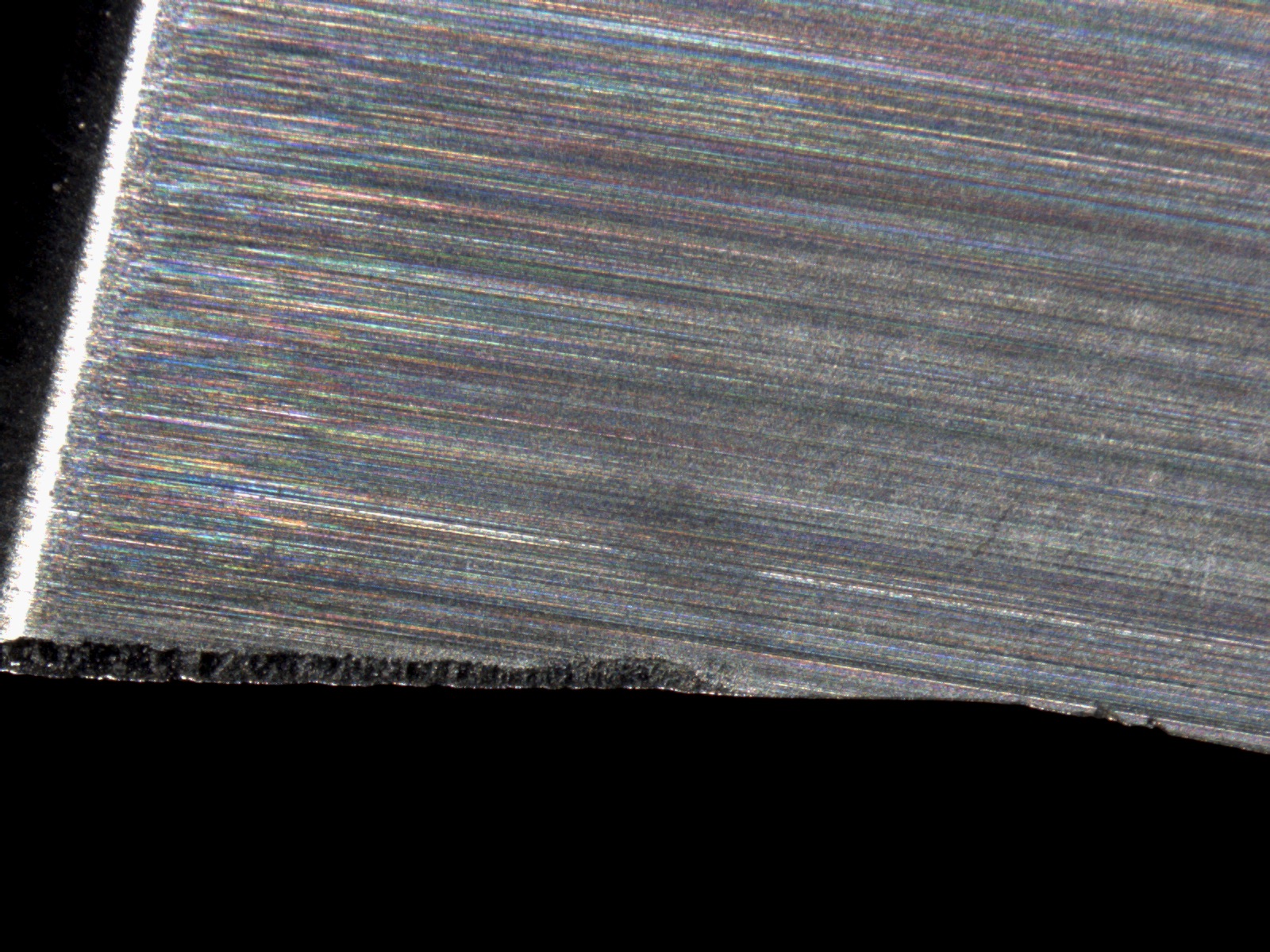}
	\caption{Example of combination of flank wear (left to middle) and chipping (right).}
	\label{fig: VBAndChipping}       
\end{figure}

\begin{center}
\begin{table}
  \begin{tabular}{ | l | c | c |}
    \hline
     \textbf{\makecell{Wear mechanism}} & \textbf{\makecell{Frequency (relative)}} \\ \hline
    Flank wear & 536 (82.72\%)  \\ \hline
    Chipping & 359 (55.40\%) \\ \hline
    No wear & 96 (14.81\%) \\ \hline
    Built-up edge \cite{Black1995} & 90 (13.89\%) \\ \hline
  \end{tabular}
  \caption{Frequency of wear mechanisms in our dataset.}
\label{Table: FrequencyOfWearMechanisms}
\end{table}
\end{center}
\vspace{-20pt}
\hspace{\parindent}
 Regarding the data, it is important to understand that the images depicted in figure \ref{fig: VB_example} and \ref{fig: Chipping_example} are abnormally easy cases compared to the rest of the data set. Figure \ref{fig: VBAndChipping} shows a more representative image: both flank wear and chipping are present and the areas of chipping are relatively small.
 
\subsection{Suggestion}
\label{subsec: Suggestion}
In this subsection, we address the classification tasks described above and explain the selection of certain options for the machine learning approach.

In general, deep neural networks require a large amount of data (e.g. millions of samples) to be trained \cite{Oquab2014}. However, in our case, relatively little data is available. For such cases, transfer learning  has proven to be successful for other image classification tasks \cite{Oquab2014, Shin2016}. Transfer learning with deep neural networks refers to reusing the first part of a network which has been trained on a different task with a big data set. To be precise, one uses the already trained weights of the first layers of a neural network. These first layers perform feature extraction. Research has shown that these learned features can often be successfully transferred from one task to another \cite{Oquab2014, Shin2016}. 

Consequently,  we apply transfer learning in our first cycle. First, we use an already trained deep neural network to extract features from our images. Since this outputs a large quantity of features we need a feature selection mechanism. Thus, we apply a gradient boosting classifier that automatically performs feature selection as part of the classification \cite{Friedman2011}. To be precise, we apply the gradient boosting classifier to perform the classification regarding chipping and flank wear.

As evaluation strategy, we choose 3-fold cross-validation. Cross-validation is applied to use the whole data set and since it gives a good estimate for the error on unseen data \cite[p. 241]{Friedman2011}.  We choose only three folds due to the high run time of the algorithms. As evaluation metric, we report the Matthews Correlation Coefficient (MCC) \cite{Matthews1975} since our dataset is imbalanced for both classification tasks. The MCC takes class imbalance into account --- a MCC of ``0'' corresponds to random guessing based on the relative size of the classes. Perfect predictions yield an MCC of ``1'', ``-1'' indicates that the predictions are inverse to the actual labels. Also, contrary to other popular evaluation measures like Precision, Recall and F-Measure the MCC also takes the true negatives into account. Thus, it gives a more holistic assessment of the classifier's performance \cite{Powers2007}. Additionally, we report the confusion matrix to enable a more in-depth evaluation of the different types of correct and false predictions.

\subsection{Development}
In this subsection, we describe the implementation of our approach in the Python programming language.

\begin{figure}[thb]
    \centering
	\includegraphics[width=1.0\linewidth]{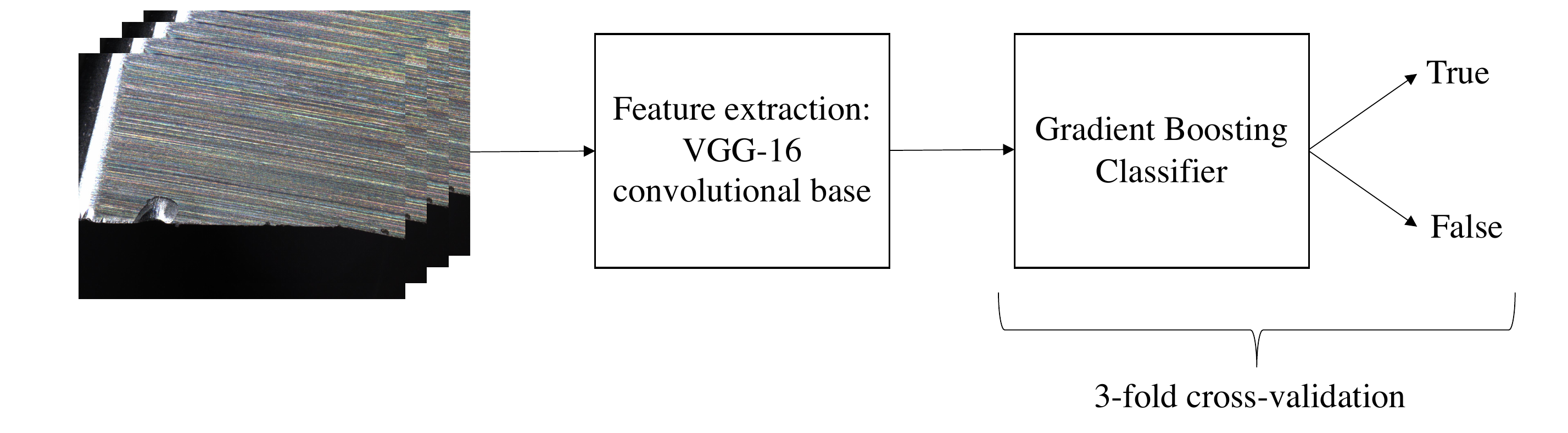}
	\caption{Overview of classification pipeline.}
	\label{fig: Pipeline}  
\end{figure}

The images in our dataset have a size of 1600x1200 pixels. Since so far our computations are performed on a standard laptop we resized the images to 640x480 pixels to speed up computation. For the classification, we apply the transfer learning approach described in section \ref{subsec: Suggestion}. Figure \ref{fig: Pipeline} shows an overview of the pipeline. We use the convolutional base of a VGG-16 network \cite{Simonyan2015} to extract features from the raw images. The convolutional base comprises the first layers of a convolutional neural network, i.e. all layers apart from the fully-connected ones  and the last softmax layer. The VGG-16 network which we apply, is pretrained on the  \texttt{ImageNet}
dataset \cite{JiaDeng2009}. Specifically, we use the implementation from the \texttt{Keras}
package \cite{chollet2015keras}. At this point, the 153,600 features per image are saved to disk since the computation of these features is time consuming and independent of the concrete classification task.
These features are then used as input for two classification models. One with chipping  and one with flank wear as binary target variable. The gradient boosting classifier implemented in the \texttt{scikit-learn} package \cite{scikit-learn} is applied for these classification tasks. 

\subsection{Evaluation}
After describing the implementation of the classification pipeline, we now present the results of the two classification models in terms  of  Matthews Correlation Coefficient and confusion matrix.

The Matthews Correlation Coefficient resulting from the flank wear classifier is 0.878. Table \ref{ConfMat flank wear} contains the corresponding confusion matrix.

For the chipping classifier the Matthews Correlation Coefficient is 0.644. The corresponding confusion matrix is shown in table \ref{ConfMat chipping}.
\begin{center}
\begin{table}
  \begin{tabular}{ | l | c | c |}
    \hline
     & \textbf{\makecell{Predicted\\flank wear}} & \textbf{\makecell{Predicted\\no flank wear}} \\ \hline
    \textbf{\makecell{Actually \\flank wear}} & 532 & 4 \\ \hline
    \textbf{\makecell{Actually \\no flank wear}} & 18 & 94 \\ \hline
  \end{tabular}
  \caption{Confusion matrix for the flank wear classifier.}
\label{ConfMat flank wear}
\end{table}
\end{center}
\begin{center}
\begin{table}
  \begin{tabular}{ | l | c | c |}
    \hline
     &\textbf{\makecell{Predicted\\chipping}} & \textbf{\makecell{Predicted\\no chipping}} \\ \hline
    \textbf{Actually chipping} & 292 & 67 \\ \hline
    \textbf{Actually no chipping} & 48 & 241 \\ \hline
  \end{tabular}
\caption{Confusion matrix for the chipping classifier.}
\label{ConfMat chipping}
 \end{table}
\end{center}
\vspace{-35pt}
\subsection{Conclusion}
Our results show that it is possible to use deep learning to extract relevant features and perform classification regarding wear mechanisms based on our raw images. Keeping in mind, that a  Matthews  Correlation Coefficient  of ``0'' corresponds to random guessing based on the class sizes our results are significantly better. Discussions with domain experts confirmed that the approach is promising. The usefulness, however, can be increased when the location and extent of wear mechanisms are determined as well.
\section{Future cycles: Semantic Segmentation and Business Impact \& Usability}
\label{FutureCycles}
In this section we present the currently planned second and third cycle on a conceptual level. For easier reading we refrain from using the dedicated steps in the DSR cycle (awareness, suggestion etc.) in this section. 
\subsection{Second Design Cycle: Semantic Segmentation}
\vspace{-5pt}
In the first design cycle, we have shown that it is possible to use deep learning to extract relevant features for a classification regarding wear mechanisms based on our raw images. Discussing the results with domain experts, we learned that a more detailed characterization of images from worn tools would be beneficial. In detail, an exact identification of the location as well as the extent of wear phenomena would significantly increase the impact of our system. First, this enables statistics over certain wear phenomena. For instance, flank wear is a widespread tool life characteristic---the corresponding ISO Norm for turning \cite{ISO3685} recommends 0.3 mm as tool life criterion. Measurements of flank wear on many tools from one process give an indication if the tools are changed too late, too early or just right. Second, heatmaps can be generated which show the locations of frequent wear. Accordingly, our research question for the second cycle is: \textbf{How can we design a system for deep-learning-based computer vision to automatically determine the location and extent of wear phenomena on images from worn tools?}

\begin{figure}[!tbp]
  \centering
  \subfloat[Raw image.]{\includegraphics[width=0.23\textwidth]{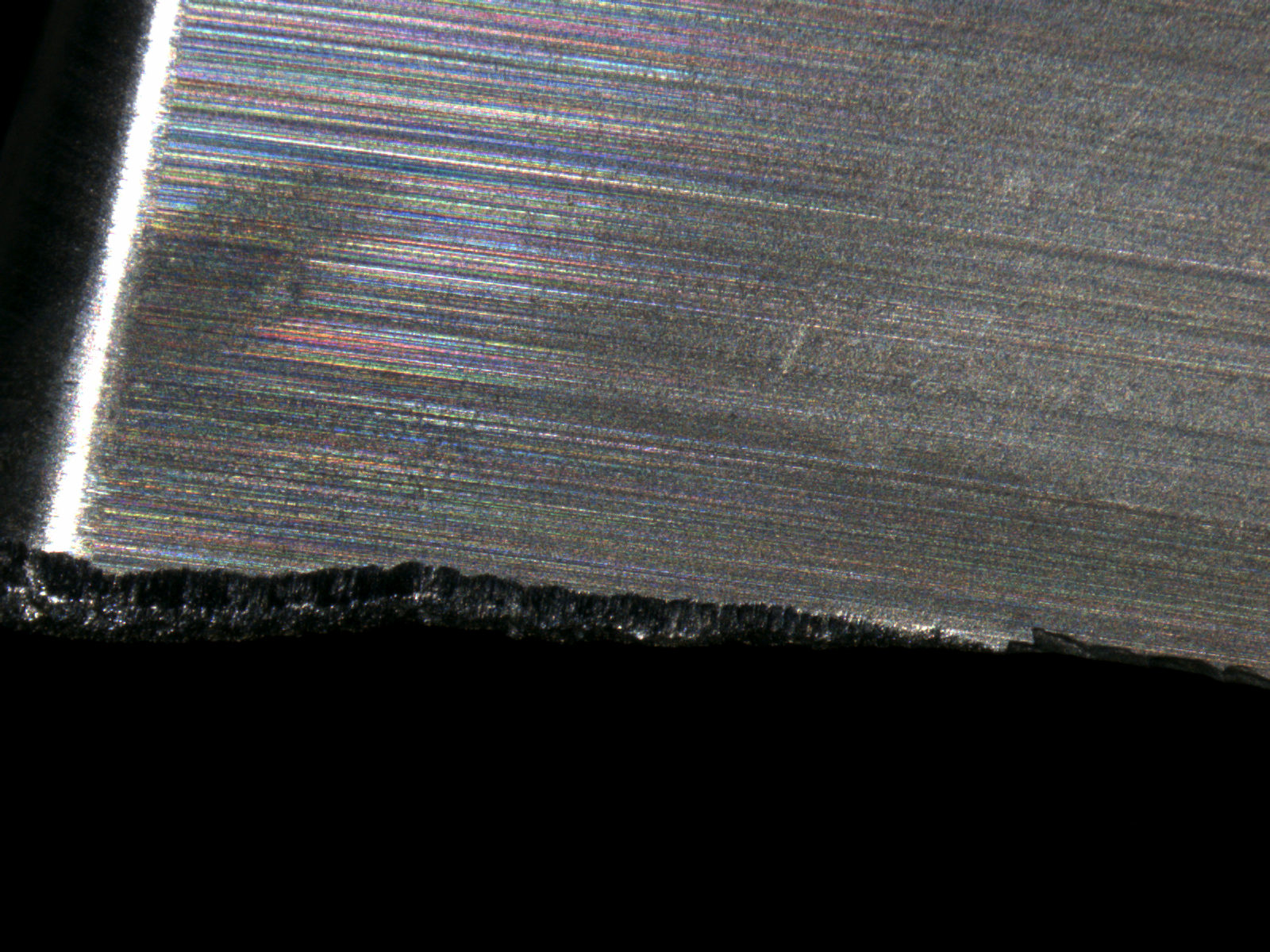}\label{fig:f1}}
  \hfill
  \subfloat[Segmentation map: flank wear in light grey and chipping in white.]{\includegraphics[width=0.23\textwidth]{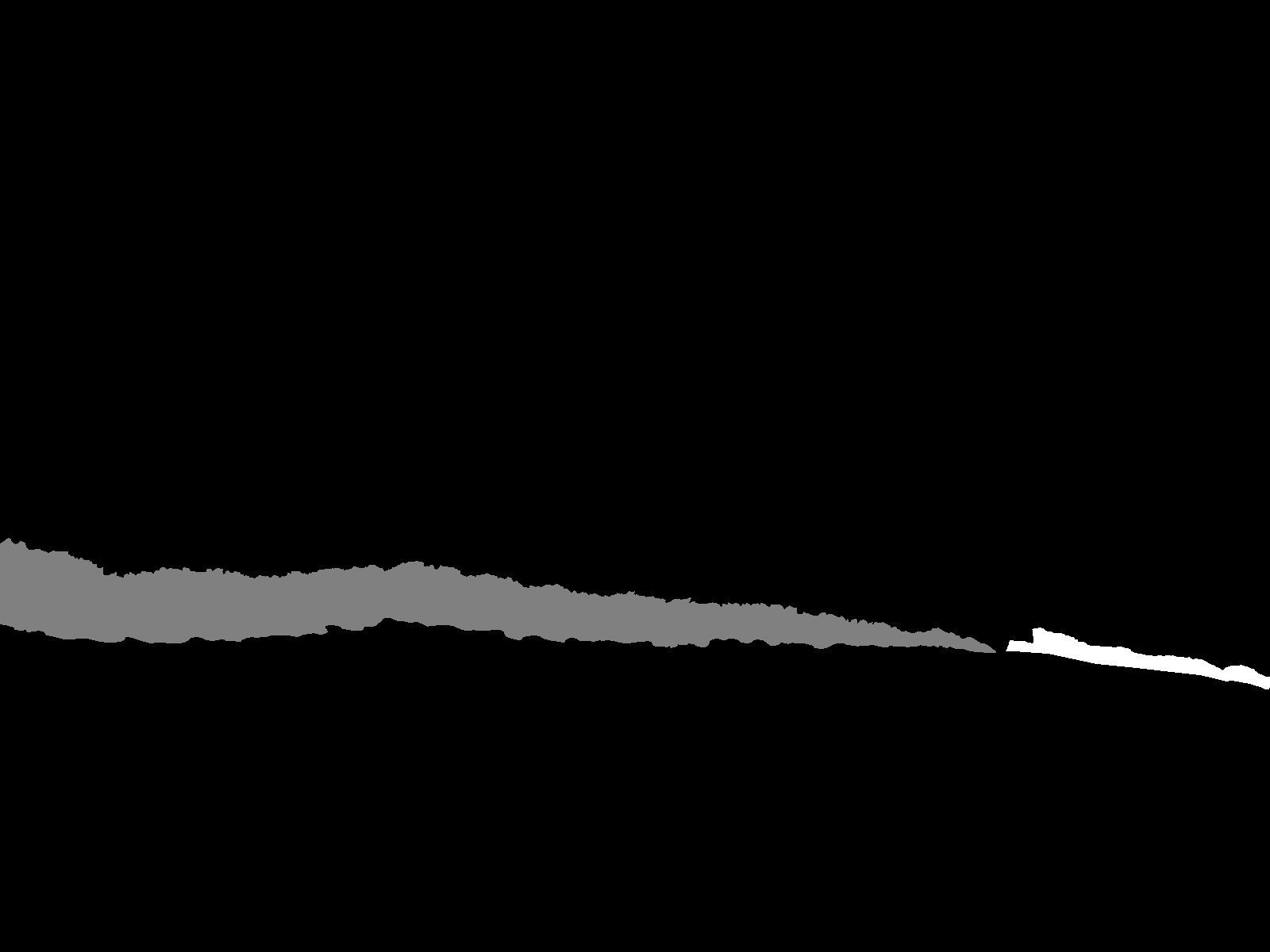}\label{fig:f2}}
  \caption{Example of raw image and corresponding segmentation map.}
  \label{fig:RawAndSegmentationMap}
\end{figure}

In the following, we present how we propose to address this research question in the future. To extract the location and extent of wear mechanisms from the images, we propose a system for automatic semantic segmentation. The goal of semantic segmentation is to classify each pixel in a given image into a fixed set of categories \cite{he2017}. Figure \ref{fig:RawAndSegmentationMap} illustrates this by depicting an original image and the corresponding labels. So far, we are manually generating this pixelwise labelling as input for supervised learning. The goal of the second cycle is to automatically generate such labels.

Research shows that deep convolutional neural networks are the best known approach for semantic segmentation; compare Ronneberger et al. (2015) \cite{Ronneberger2015} for the U-Net architecture and Long et al. (2015) \cite{Long2015} for the so-called Fully Convolutional Networks for Semantic Segmentation. Accordingly, we will implement our system for the second cycle based on these network architectures. Since the pixelwise labelling is labor-intensive, we will explore if data augmentation techniques can help to reduce the number of required labelled images. Data augmentation refers to generating more training images by applying operations like shifting, rotating, flipping, distorting etc. to the original images \cite{Perez2017}. Previous research shows mixed results: Long et al. (2015) \cite{Long2015} note that data augmentation does not help for their task, while Ronneberger et al. (2015) \cite{Ronneberger2015} describe data augmentation as an essential part of their approach. 

We plan to evaluate this system for automatic semantic segmentation as follows. Training deep neural networks involves the optimization of hyperparameters such as the learning rate. Accordingly, we will split our data into three disjoint sets. The training set is used to learn the weights of the neural network, the validation set is used to find optimal hyperparameters, and the test set gives an estimate for performance on unseen data \cite{Goodfellow2016}. 

Since our images contain a lot of background and unworn tool surface the choice of a proper evaluation measure is crucial. A popular and well-suited choice is the Intersection over Union (IoU) \cite{RahmanAtiqur2016}. It is defined as 
\begin{center}
    $IoU=\frac{True\: Positives}{True\: Positives+False\: Positives+False\: Negatives}$.    
\end{center}Thus, the intersection between the labelled area and the area predicted by the algorithm is divided by the union of these two areas, hence the name. Depending on the use case this measure can then be aggregated, e.g., over all pixels or over different classes of wear mechanisms. 

\subsection{Third Design Cycle: Business Impact \& Usability}
\vspace{-5pt}
Whilst the first two design cycles focus on technical feasibility, implementation and statistical performance, the third cycle will focus on business impact and usability. Thus, the research question we address in the third cycle is: \textbf{How is the business impact and usability of the system for semantic segmentation perceived by the users?}

In order to investigate this research question, we envision to examine two different scenarios which we describe in the following: First, the application of the system to improve process optimization. Second, the application of the system to optimize tool development.

Currently, customers of a tool manufacturer request an inspection of an application engineer in case they see optimization potential regarding their production process. Then the application engineer visits their production line and works on optimizing the production process.
Thus, usually he\footnote{To ensure a steady reading flow in this work, we use only one gender and use male pronouns (he, his, him) when necessary. This always includes the female gender as well.} just looks at a small number of worn tools which are obtained during his visit or shortly before. Our proposed system enables an improved scenario: Again, a customer assuming optimization potential in a machining process requests a visit of an application engineer. He is then asked to collect all worn tools from the respective process for the next days/weeks. These worn tools are then sent to the application engineers and automatically analyzed by our proposed system. This has two major advantages. First, the application engineer receives the results of the wear characterization already before visiting the production line. This enables him to prepare better and to focus on the actual problem to be solved. Second, he gets deeper insights since the sample of worn tools is bigger and more representative which is even more important than the first advantage.

Thus, application engineers of a tool manufacturer are an important user group of our proposed system. Of course, we will also consider the customers of the application engineers. To ensure real-world impact, we will assess the business impact and usability of our proposed system in a field experiment: application engineers use the system for a certain time in their daily work. Afterwards, we interview both the application engineers and their customers regarding the business impact and usability of our proposed system.

Supporting tool development is another promising application of the system for semantic segmentation. When used with images from many different customers the proposed system can be utilized to understand inherent problems of certain tools and needs of the market. For example, if a certain tool suffers from severe chipping even though utilized at different customers on different material and with different process parameters, this is an important indication for the next generation of tools. Such an analysis can be another example for successful value co-creation: the development of tools tailored to the most prevalent problems in the market is only possible when information is shared between tool users and manufacturer.
This is particularly promising since according to domain experts there is relatively little communication between tool manufacturers and companies using the tools. The proposed system could alleviate this problem. Often, the exchange of data or information between companies is restricted due to data confidentiality concerns \cite{Hirt2018}. Worn tools are already sold from the companies using them to special recycling companies, thus they are not considered as confidential information. 

Consequently, tool developers are another important user group of our proposed system. We will perform a field experiment with this group to ensure real-world impact: they use our proposed system for a certain time and then we will interview them regarding the business impact and usability.

\section{Conclusion}
\label{conclusion}
\vspace{-5pt}
So far, research regarding the management of products having reached their end-of-life focuses on facilitating sustainable solutions like refurbishing and recycling instead of e.g., landfilling. In the work at hand, we propose an approach to generate additional value from products having reached their end-of-life. An exemplary use case in the machining industry illustrates how an automatic characterization of worn tools can foster value co-creation between tool manufacturer and the users of the tool. Both parties can benefit from easier and better process optimization and tool development. 

There are four main contributions of this work: First, we summarize the state-of-the-art in automatic wear characterization on machining tools and show how such systems can be used beneficially apart from tool condition monitoring. Second, we show the feasibility of a deep-learning-based classification approach for different wear phenomena. With first results at hand, we, thirdly, present our agenda for future research. From a technical point of view, it will enable a complete characterization of worn tools including details like the exact location and extent of each wear phenomena. From a business point of view, it will evaluate the actual impact of the system. As fourth and more general contribution, we describe an example of how deep learning and products which have reached their end-of-life can be leveraged to positively impact earlier stages of the value chain. Thus, we argue that in certain cases products having reached their end-of-life should be considered an asset. This approach can be promising for further applications: experts in the respective domain confirm the potential usefulness of analyzing worn industrial seals. Furthermore, several applications in the Business-to-Consumer setting seem feasible: for instance, worn shoes could be analyzed to improve future generations of shoes.

Besides these contributions, this work has limitations. On a general level, parts of this paper are still conceptual. A more specific limitation regarding the process optimization and tool development use cases is that the survivorship bias \cite{Brown1992} has to be kept in mind: in extreme cases, machining tools can break completely and customers will (probably) not send back these tools. Consequently, our proposed system cannot generate a complete overview of the wear mechanisms in real production processes. Another technical limitation is our (so far) limited consideration of only the flank of a worn cutting edge. Analyzing the other side (called face \cite{ISO3685}) can also provide valuable information. However, domain experts confirmed the usefulness of an automatic characterization of the flank side. Thus, we believe this is a reasonable scope for now and leave this aspect for future work.

Overall, we believe this ``cutting edge'' research is a promising field of research. It has potential real-world impact and extends the research on value co-creation by showing possibilities based on forensic analyses of products having reached their end-of-life.



\bibliographystyle{ieeetr}
\bibliography{references}

\end{document}